\documentclass[aps,twocolumn,prl,superscriptaddress,floatfix,showpacs]{revtex4}
\usepackage{graphicx}

\newcommand{\bi}{\bibitem}
\newcommand{\ba}{\begin{eqnarray}}
\newcommand{\ea}{\end{eqnarray}}

\def\ct{\cite}
\def \beq{\begin{equation}}
\def \eeq{\end{equation}}
\def \bea{\begin{eqnarray}}
\def \eea{\end{eqnarray}}

\begin{document}

\title{Quench Dynamics of Edge States in 2-D Topological Insulator  Ribbons}
\author{Aavishkar A. Patel}
\author{Shraddha Sharma}
\author{Amit Dutta}
\affiliation{Department of Physics, Indian Institute of Technology Kanpur, Kanpur 208016, India}

\begin{abstract}
We study the dynamics of edge states of the two dimensional BHZ Hamiltonian in a ribbon geometry following a sudden quench to the quantum critical point separating the topological insulator phase from the trivial insulator phase. The effective edge state Hamiltonian is a collection of decoupled qubit-like two-level systems which get coupled to bulk states following the quench. We notice a pronounced collapse and revival of the Loschmidt echo for low-energy edge states illustrating the oscillation of the state between the two edges. We also observe a similar collapse and revival in the spin Hall current carried by these edge states, leading to a persistence of its time-averaged value.
\end{abstract}

\pacs{64.70.Tg, 03.65.Pm, 74.40.Kb}

\maketitle

Topological insulators (TIs) are novel materials with an insulating bulk and conducting edges which are of extensive contemporary interest~\cite{kane05,bernevig06,hasan10,qi11}. The low-energy electrons in two dimensional (2-D) Hg-Te/Cd-Te quantum well TIs, which display conducting helical edge state solutions that exist within the bulk bandgap in the TI phase, are described by the 2-D BHZ Hamiltonian~\cite{bernevig06,qi11}. The bulk states undergo a quantum phase transition (QPT)~\cite{sachdev99,chakrabarti96,sondhi97} with the low-energy modes satisfying a 2-D Dirac Hamiltonian (DH) with a linear dispersion at the quantum critical point (QCP) (which is a 2-D Dirac point). Additionally, the chiral edge states with linear dispersion in the TI phase are described by an effective 1-D DH~\cite{bernevig06}.

At the same time, there is a recent upsurge in studies  of quenching dynamics of quantum many body systems across QCPs~\cite{polkovnikovrmp,duttarmp10,dziarmaga10}, essentially because of the possibility  of experimental realization of the same in optical lattices ~\cite{greiner02}. The scaling of the defect density generated in the final state following a slow~\cite{zurek05,damski05} or a sudden quench~\cite{grandi10,mukherjee11}, or generation of quantum correlations which are otherwise absent in the defect free final state~\cite{sengupta09} or the possibility thermalization with an effective temperature~\cite{rigol08} are some of the topics which are being explored thoroughly.

In this communication, we focus on the dynamics of edge states of the BHZ Hamiltonian when the system is suddenly quenched from the TI phase to either the QCP or the trivial insulator (TrI) phase. The question here is whether there is a surviving edge current following the quench. (It is to be noted that when a one-dimensional chain of hard core Bosons is quenched from the superfluid to the Mott Insulator state, there is a surviving supercurrent in the insulator phase which oscillates in time~\cite{klich07}). In our problem, the quench couples the two-level subspace of the edge states to a multi-level environment of bulk states. We study the decoherence of these edge states using the Loschmidt echo (LE)~\cite{quan06} when the 2-D Hamiltonian is quenched from the TI phase to the QCP (or to the TrI phase). We observe a strong oscillation of the low-energy edge states between the two edges of the system and the time-averaged persistence of the spin Hall current (SHC) when the system is quenched to the QCP, which we attribute directly to the linear low-energy dispersion at this point (which is also found in other models describing 2-D TIs). The experimental prospect of real-time tuning of parameters controlling QPTs of TIs in optical~\cite{stanescu09} and photonic lattices~\cite{rechtsman12} as well as by exploiting Floquet dynamics~\cite{lindner11}, has made the study of quenching dynamics of TI Hamiltonians relevant and important. It is to be noted that slow quenching results in a violation of the Kibble-Zurek scaling of defects in systems with edge states~\cite{bermudez10}.

The  $4 \times 4$ BHZ Hamiltonian comprising of two 2$\times$2 blocks (for opposite electron spins) is given by
\begin{equation}
\mathcal H_{BHZ} = \left(
\begin{array}{clrr} 
    H(\vec{k}) &0 \\ 
    0  &H^*(-\vec{k})
\end{array}
\right),
\label{Eq:Ham_BHZ}
\end{equation}
where $H(k)=[C-D(k_x^2+k_y^2)]{\bf I}_{2\times2}  + A[k_x\sigma^{x}+k_y\sigma^{y}] + [m-B(k_x^2+k_y^2)]\sigma^{z}$. Here, $A,B,C,D$ and $m$ are determined by the thickness of the quantum well and the material parameters; the parameter $m$ controls the phase of the system and changes sign relative to $B$ when the system crosses from the TI phase (where edge states are present) to the TrI phase (with no edge states) via a DP at $m/B=0$. Although the results presented here are valid in generic situations, we shall set $D=0$ for simplicity, which also ensures an electron-hole symmetric spectrum. We consider a ribbon geometry extending from $-L/2$ to $L/2$ in the $y$ direction (with the wavefunction vanishing at the edges) and apply periodic boundary conditions in the $x$-direction~\ct{zhou08,konig08,wada09}.

\begin{figure}[h]
\includegraphics[width=7.0cm]{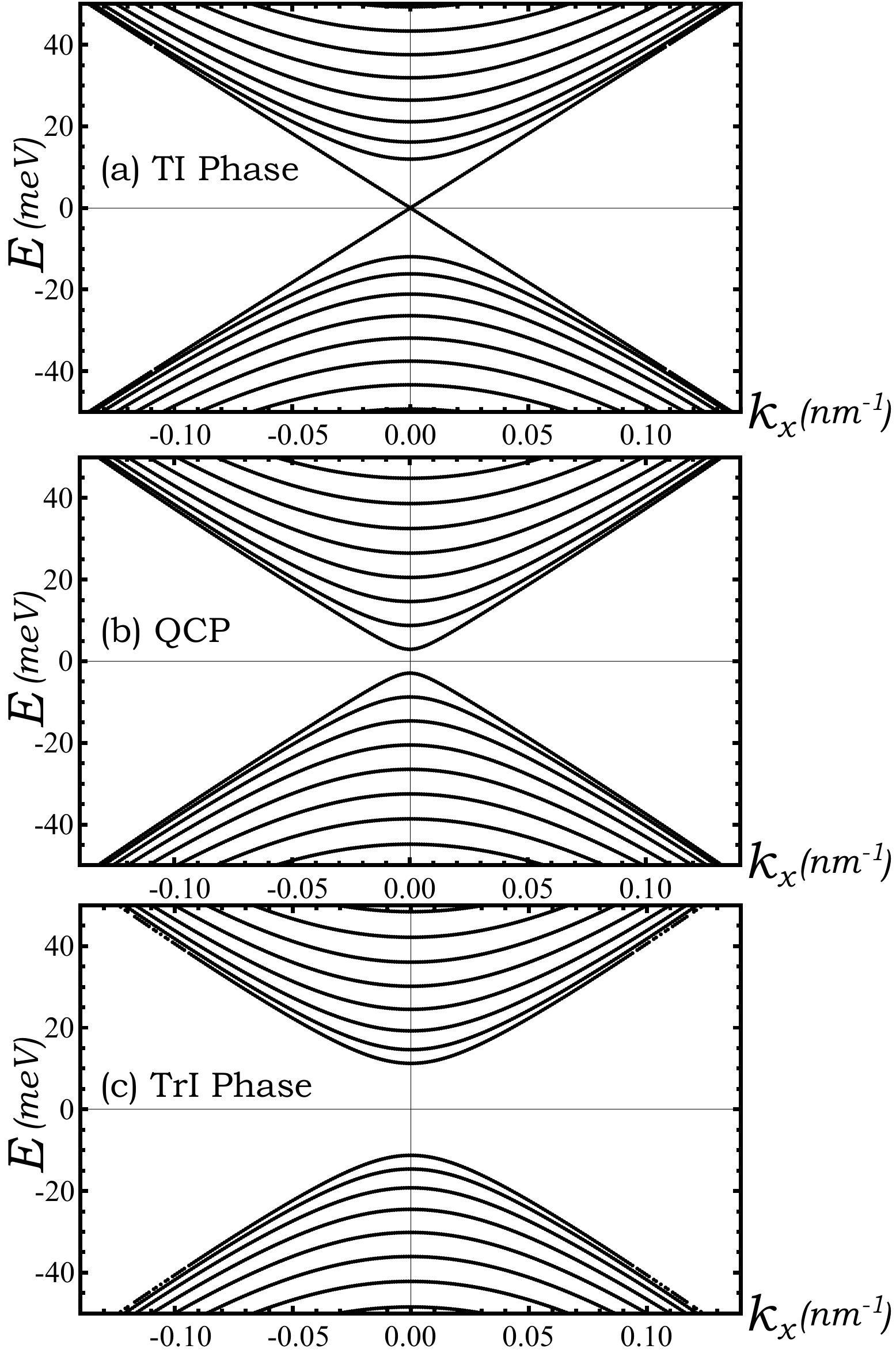}
\caption{Spectrum of the BHZ Hamiltonian in a ribbon geometry with ribbon width $L=200~nm$. The parameters used are $A=364.5~meV/nm$, $B=-686~meV/nm^2$ and $C=D=0$. The values of $m$ used are $-10~meV$ (a), $0$ (b) and $+10~meV$ (c). There is a small gap of $O(A/L)$ in the spectrum near $E=0$ at $m=0$ because of the finite width of the ribbon. There also exists an exponentially small gap between the edge state bands in the TI phase. Note the (almost) equal spacing of energy levels for $k_x=0$ at $m=0$, characterized by the solutions of a Dirac particle geometrically confined in a 1-D box.}
\label{Fig:Spectrum}
\end{figure}

To obtain the spectrum of Hamiltonian~(\ref{Eq:Ham_BHZ}), we consider the $2\times2$ block $H(k)$, fix $k_x$, and use $k_y\rightarrow -i\partial_y$, with the trial solution $\psi=(A_1,A_2)e^{\lambda y}$~\cite{zhou08}. The condition that the wavefunctions must vanish at the edges of the ribbon quantizes the energies $E$ of the eigenstates at a given $k_x$, which are given by the solutions of the following transcendental equation:
\begin{eqnarray}
\frac{\tanh(\lambda_+L/2)}{\tanh(\lambda_-L/2)}+\frac{\tanh(\lambda_-L/2)}{\tanh(\lambda_+L/2)}=~~~~~~~~~~~~~~~~~~~~~~~~~~~&& \nonumber \\
\frac{\lambda_+^2+\lambda_-^2-(B/A)^2(\lambda_+^2-\lambda_-^2)^2}{\lambda_+\lambda_-},~~~~~~&&
\end{eqnarray}
where $\lambda_{\pm}=\sqrt{k_x^2+F  \pm \sqrt{F^2-(M^2-E^2)/B^2}}$, with $F=(A^2-2MB)/(2B^2)$. The corresponding wavefunctions are given by
\begin{equation}
\psi(x,y)=e^{i k_x x}\left(c_{+}f_{+}(y)+c_{-}f_{-}(y)\right),
\end{equation} 
where $c_{\pm}$ are two-component spinors whose entries are determined by the boundary conditions, with $f_+(y)=\cosh(y\lambda_+)/\cosh(L\lambda_+/2)$$-\cosh(y\lambda_-)/
\cosh(L\lambda_-/2)$, and $f_-(y) =\sinh(y\lambda_+)/\sinh(L\lambda_+/2)$$-\sinh(y\lambda_-)/\sinh(L\lambda_-$
$/2)$. 

One can show that  in the TI  phase ($m/B>0$), there are two types of eigenstates of the Hamiltonian, i.e., edge states (localized towards the edges and decaying exponentially over a length $1/\lambda_-(k_x,m)$) and  bulk states (spreading across the whole ribbon). In order to have true edge states, one needs $L\gg1/\lambda_-$. The spectrum, which is symmetric in $\pm k_x$ and $\pm E$,  is displayed in Fig.~\ref{Fig:Spectrum}. The edge states in the TI phase exist for $|k_x|<k_0$, $k_0$ depends on $m$ and $L$~\ct{konig08}. The solutions of the two $2\times2$ blocks are time reversed conjugates of each other, with the same set of energies but opposite momentum and spin.

We now perform a sudden quench of the parameter $m$ going from $m/B >0$ to  $m/B \leq 0$ and look at the subsequent evolution of an edge state and its spin current. The edge states, which originally formed a qubit-like two level system with an effective Hamiltonian $H_{edge} \approx A k_x \sigma_z$ at each $k_x$~\cite{qi11} now get coupled to several bulk modes and subsequently decohere. Following a sudden quench, the evolution of an edge state is given by
\begin{equation}
|\psi_{edge}(k_x,t)\rangle = \sum_{n=-\infty}^{\infty} \langle\psi_{n}(k_x)|\psi_{edge}(k_x)\rangle e^{-iE_nt} |\psi_{n}(k_x)\rangle,
\label{Eq:Evolution}
\end{equation}
where $|\psi_{edge}(k_x)\rangle$ is an edge eigenstate of the Hamiltonian at the initial value of $m$ ($=m_1$) and $|\psi_{n}(k_x)\rangle$ are the eigenstates of the Hamiltonian at the final value $m_2$. The index $n$ runs from $-\infty$ to $\infty$ excluding $n=0$ and denotes the -ve and +ve energy bulk modes, respectively. Since all the modes are plane waves along the $x$ direction, different $k_x$ modes do not couple to each other. To study the dynamics of a single edge state and quantify its decay, we calculate the LE $\mathcal{L}(t)=|\langle\psi_{edge}|e^{iH(m_1)t}e^{-iH(m_2)t}|\psi_{edge}\rangle|^2$, which using Eq.(\ref{Eq:Evolution}) can be put in the form
\begin{equation}
\mathcal{L}_{edge}(k_x,t) = \left|\sum_{n=-\infty}^{\infty} |\langle\psi_{n}(k_x)|\psi_{edge}(k_x)\rangle|^2 e^{-iE_nt}\right|^2.
\label{Eq:LE_Fourier}
\end{equation} 

\begin{figure}[h]
\includegraphics[width=7.75cm]{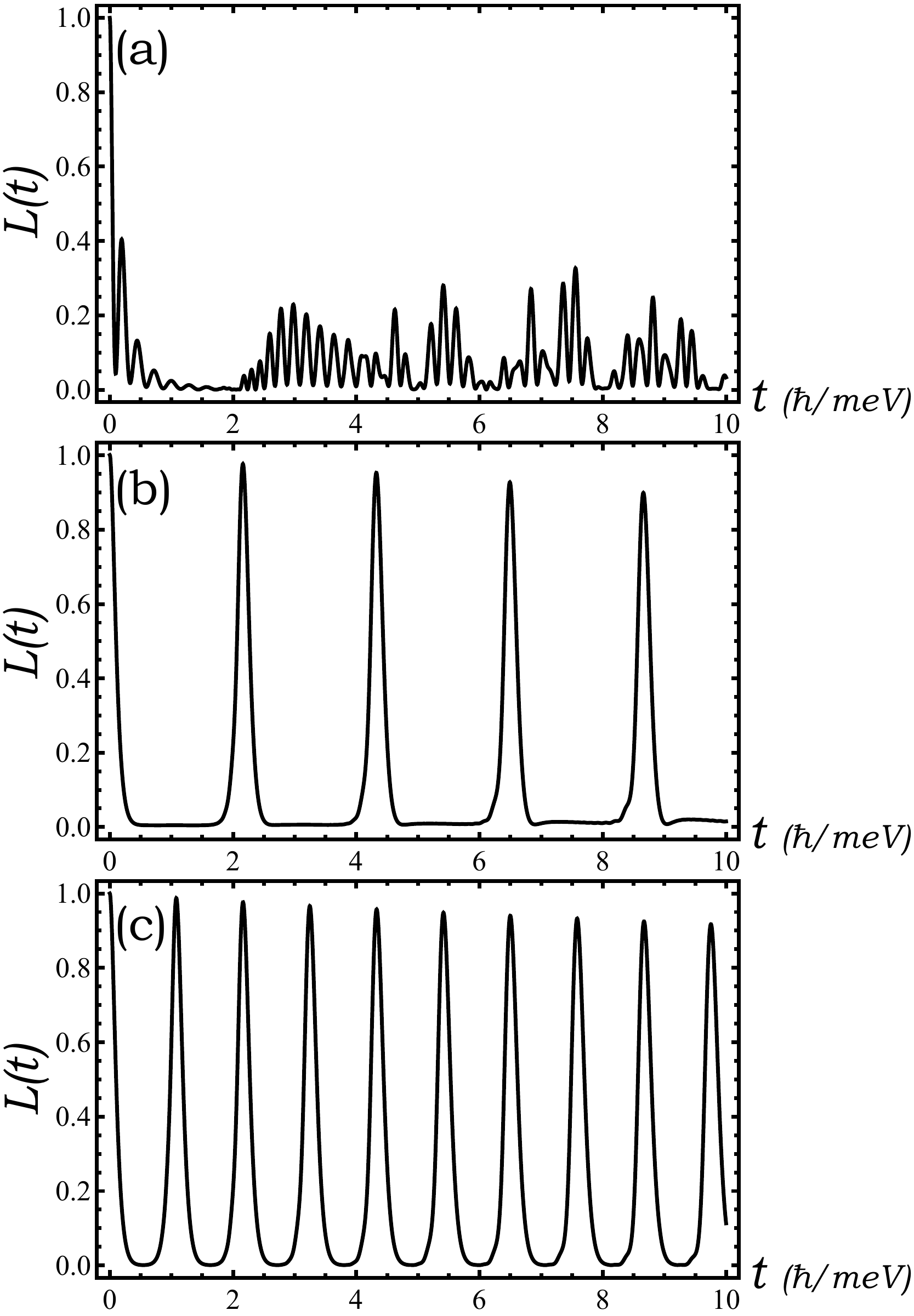}
\caption{Loschmidt Echo for various edge states and quenches. The system parameters are $A=364.5~meV/nm$, $B=-686~meV/nm^2$, $C=D=0$ and $L=400~nm$. LE for an edge state with: (a)  $k_x=0.01~meV/nm$ and $m=-10~meV$ after quenching to $m=+10~meV$. There is no significant revival of the edge state. (b)  $k_x=0.001~meV/nm$ and $m=-10~meV$ after quenching to the QCP at $m=0$. There is a pronounced collapse to $0$ and a nearly full recovery of the LE for several cycles. (c) $k_x=0$ and $m=-10~meV$ after quenching to the QCP. There is a doubling of the frequency of oscillation as compared to the previous case, as this edge state exists on both edges.}
\label{Fig:Edgestate_Echo}
\end{figure}

In general, the LE defined above  initially drops rapidly with time and turns into a rapidly oscillating noisy function of small amplitude (Fig.~\ref{Fig:Edgestate_Echo}(a)), indicating that the edge state decoheres significantly. However, there is a striking difference when one looks at the evolution of a low-energy ($k_x<<k_0$) edge state following a quench to the QCP at $m=0$;  the LE of the edge state shows a pronounced collapse and nearly complete revival for several oscillation cycles (Fig.~\ref{Fig:Edgestate_Echo}(b)). This is  a consequence of the nearly equal spacing of the first few energy levels at low $k_x$ near $m=0$ (arising due to confinement of the linearly dispersing particles ($\nu=1$ at the QCP) in a ribbon geometry)  where the overlap with the edge state is the most significant (see Fig.~\ref{Fig:Spectrum}).
We then have  $E_n\approx\mathrm{sign}(n)[E_g/2+(|n|-1)\Delta E]$ for all significant terms in Eq.(\ref{Eq:Evolution}), where $E_g\sim1/L$ is the bulk bandgap. The summations  over $n>0$ and $n<0$, then represent Fourier series of a peroidic function with period 
\begin{equation}
\tau=\frac{2\pi}{\Delta E} \approx 2\hbar\frac{L}{A},
\label{eq_tau}
\end{equation} 
making the LE a periodic function with this period. Since $\mathcal{L}(t=0)=1$, the LE shows a near-complete revival at $t=n\tau$. Eventually, after several oscillations, the slight non-uniformity  in spacing becomes significant and the revival of the LE weakens. Since $\Delta E\sim1/L$, the period of this revival scales as $L$. For small quench amplitudes ($|m|<<A/L$), the edge state does not decay significantly.

\begin{figure}[h]
\includegraphics[width=7.75cm]{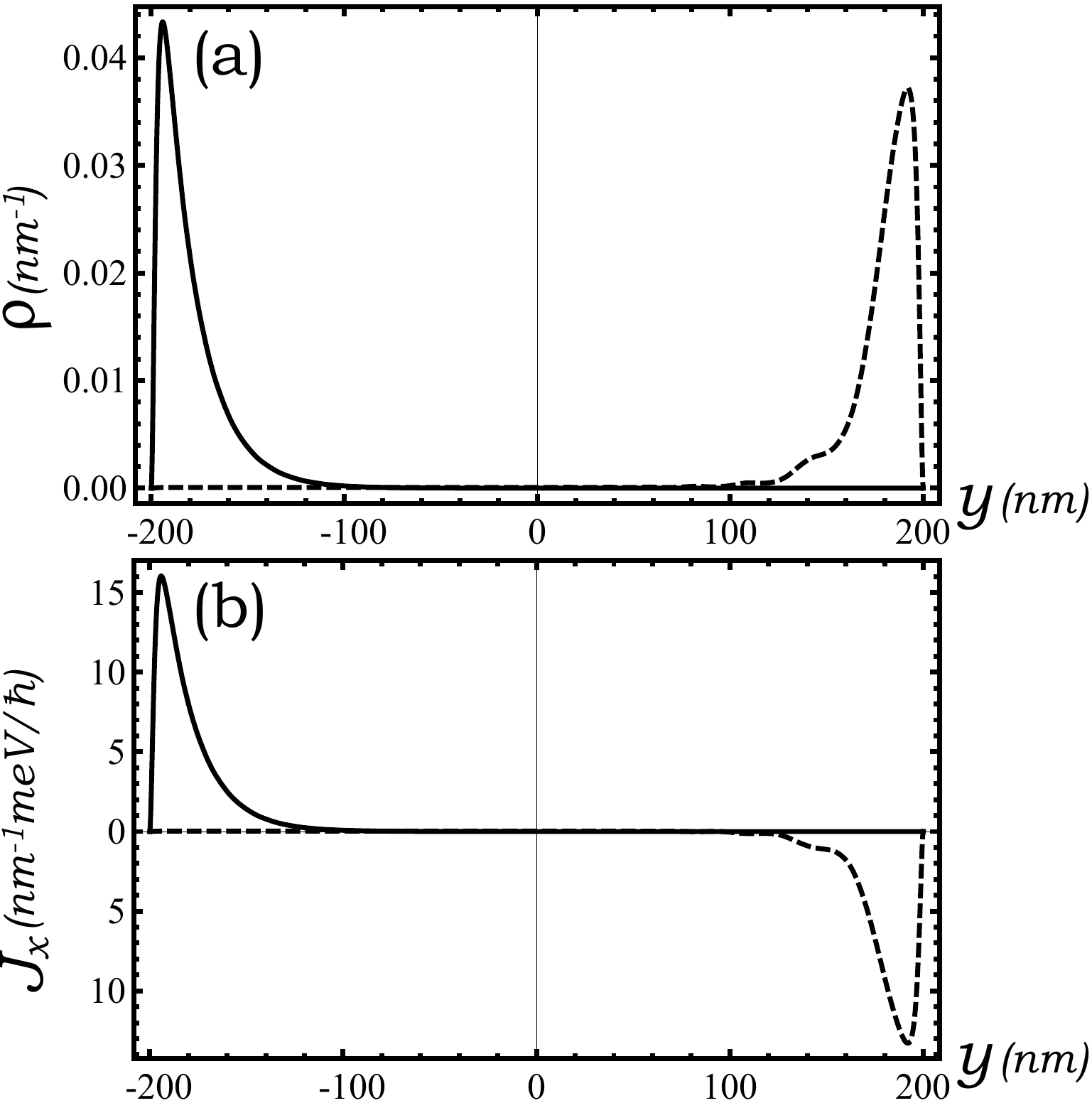}
\caption{(a) The probability density $\rho=\psi^{\dagger}\psi$ of an edge state ($k_x=0.001~meV/nm$) following a sudden quench from $m=-10~meV$ to $m=0$ shown at $t=0$ (solid) and $t=\tau/2$ (dashed). The edge state travels between the two edges. (b) The probabilty current density of the state in the $x$ direction ($J_x$) at $t=0$ (solid) and $t=\tau/2$ (dashed). The same state carries currents of opposite direction on opposite edges. The system parameters are the same as 
Fig.~(\ref{Fig:Edgestate_Echo}).}
\label{Fig:Edgestate_Travel}
\end{figure}

Interestingly, we find that the edge state travels from one edge to the other and back, existing on opposite edges at the points of maxima and minima of the LE (Fig.~\ref{Fig:Edgestate_Travel}(a)). This effect is due to the finite width of the ribbon and will not be seen in an infinite system. Since, for a significantly large system, the edge states at opposite edges do not overlap, the LE drops to zero when the edge state reaches the opposite edge, and revives again when it comes back. For very low values of momentum $k_x\rightarrow0$, the edge state exists with peaks on both edges of a finite ribbon~\cite{zhou08}. Hence, when a peak on a given edge travels to the opposite edge, the peak on the opposite edge also travels simultaneously to the given edge, resulting in a maximum of the LE at $(2n+1)\tau/2$ instead of a minimum, and hence a doubling of the frequency of oscillation of the LE. The LE is now minimum at times when the state is concenterated near the middle of the ribbon (Fig.~\ref{Fig:Edgestate_Echo}(c)).

\begin{figure}[h]
\includegraphics[width=7.75cm]{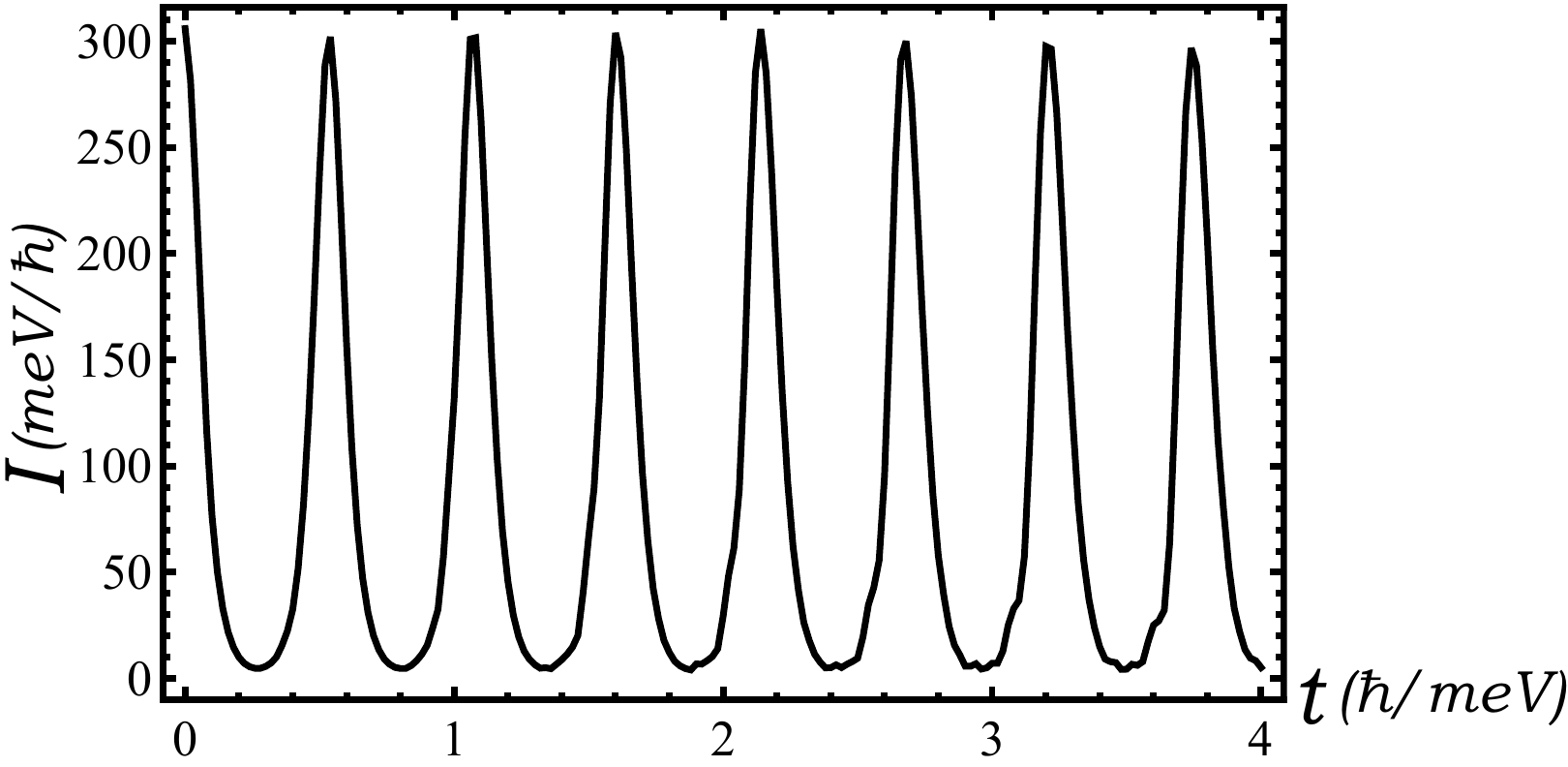}
\caption{The net probabilty current 
carried by the pair of edge states with $E=3.65~meV$ (near the $-L/2$ edge over a length $1/\lambda_-\approx37~nm$) following a quench from $m=-10~meV$ to  $m=0$. The current collapses and revives like the LE, and is always greater than zero, leading to a persistence of its time-averaged value. The system parameters are $A=364.5~meV/nm$, $B=-686~meV/nm^2$, $C=D=0$ and $L=200~nm$.}
\label{Fig:Edgecurrent}
\end{figure}

The probability current carried near the edge in the $x$ direction by the edge states over their decay length $1/\lambda_-$ is proportional to the net SHC carried by the state and its time-reversed conjugate in the opposite spin sector. It can be calculated using the continuity equation for the probability current density $\vec{J}$ in conjunction with the Schrodinger time evolution equation
\begin{eqnarray}
&&\frac{\partial}{\partial t}\left(\psi^{\dagger}(x,y,t)\psi(x,y,t)\right) + \nabla\cdot\vec{J} = 0, \nonumber \\
&&i \frac{\partial}{\partial t} \psi(x,y,t) = H \psi(x,y,t),
\end{eqnarray}
where $\psi(x,y,t)$ is the time dependent two-component wavefunction of the form  $\psi(x,y,t)=(\phi_1(y,t),\phi_2(y,t))e^{i k_x x}$. Using $\vec{k}\rightarrow-i\vec{\nabla}$, we obtain
\begin{eqnarray}
&&J_x(y,t) =2A\left(\phi_1(y,t)\phi_2^*(y,t)+\phi_1^*(y,t)\phi_2(y,t)\right) + \nonumber \\
&&2Bk_x\left(|\phi_2(y,t)|^2-|\phi_1(y,t)|^2\right),
\label{Eq:Probcurrent}
\end{eqnarray}
with $A$ and $B$ as defined in~(\ref{Eq:Ham_BHZ}). For $k_x\rightarrow0$, the first Dirac-like term dominates the second Schrodinger-like term, implying that all low-energy edge states carry virtually the same current. The profile of $J_x$ is shown at the instants of time when the edge state exists on opposite edges in Fig.~\ref{Fig:Edgestate_Travel}(b). The sign of the current reverses when the state moves to the opposite edge.

The evolution of the SHC carried by an edge state near a given edge (say, $-L/2$) can be obtained by integrating $J_x$ from $-L/2$ to $-L/2+1/\lambda_-$;  i.e.,
$I_{edge}(t)=\int_{-L/2}^{-L/2+1/\lambda} J_x(y,t) dy$.
Since there are two oppositely propagating edge states at a given energy which exist on opposite edges of the system, we must add the currents carried by both of them. The time evolution of such a current following a sudden quench to the QCP is shown in Fig.~\ref{Fig:Edgecurrent}. Due to the oscillation of the edge states between the two edges, this current also displays a pronounced collapse and revival for several cycles. The time-averaged value of the current is non-zero and is a significant fraction of the original value of the current. Thus, there is a persistence of the SHC carried by low-energy edge states following a sudden quench to the QCP. \\

After several oscillations, the edge state disperses and does not regain it's original character again. This dispersion is introduced by the non-linear $Bk^2$ term in the Hamiltonian, and must be sufficiently small if sustained oscillations are to be observed. Using the condition that the spread of the edge state over one time period must be significantly smaller than it's initial decay length $1/\lambda_-$, we obtain
the condition
${(BL)}/{A} \lesssim {1}/{\lambda_-^2}$.
Since we already have $L\gg1/\lambda_-$, this condition implies that these oscillations will be seen over an intermediate range of ribbon widths roughly determined by the values of the parameters $A$ and $B$.

It has already been mentioned that recently a realization of a Floquet TI has been experimentally achieved in photonic lattices~\cite{rechtsman12}, where the quenching dynamical study discussed above can possibly be verified. The dynamics of a Floquet quantum system driven with a time period $T$ are determined by its Floquet evolution operator $U(T)$, which may be associated with an effective Hamiltonian $H_{eff}$~\cite{kitagawa10} as $U(T)=\mathcal{T}e^{-(i/\hbar)\int_{0}^{T}H(t)dt}=e^{-(i/\hbar)H_{eff}T}$, where $\mathcal{T}$ denotes the time ordering symbol. Thus, in a Floquet system, the evolution of an edge eigenstate of $H_{eff}(m_1)$ following a sudden quench to $m_2$ is like a normal Schrodinger evolution under $H_{eff}(m_2)$ at instants of time $t=nT$. For these oscillations to be visible in experimental studies of such a system, one needs $T\ll\tau$ (see, Eq.(\ref{eq_tau})).

To summarize, we observe interesting decoherence dynamics of the low-energy edge states of a TI Hamiltonian when quenched to the QCP displayed in the temporal evolution of their LE.  Given the recent prospects of tuning of control parameters of the related Hamiltonians, we believe that this study is of experimental relevance. Furthermore, one can show  that the predicted oscillations of edge modes should occur in any 1-D or effectively 1-D system with a linearly dispersing ($\nu=1$) QCP (such as the 1-D p-wave superconducting chain with Majorana boundary modes), due to the fact that the energy levels at the QCP in a finite system will be more or less equally spaced \ct{rajak13}.

We acknowledge Apoorva Patel and Diptiman Sen for helpful discussions. AAP acknowledges the KVPY fellowship and AD and SS acknowledge CSIR, New Delhi, for financial support.

\end{document}